\documentclass[final,3p,times,twocolumn]{elsarticle}
 \biboptions{comma,sort&compress}
\usepackage{here}
 \usepackage{graphicx}

\def\nin{\noindent}
\def\beq{\begin{equation}}
\def\eeq{\end{equation}}
\def\bea{\begin{eqnarray}}
\def\eea{\end{eqnarray}}

\journal{Nuc. Phys. (Proc. Suppl.)}

\begin{document}

\begin{frontmatter}



\title{Factorization breaking in single-diffractive dijet production at the Tevatron}

 \author[label1]{Michael Klasen}
  \address[label1]{Laboratoire de Physique Subatomique et de Cosmologie,
        Universit\'e Joseph Fourier / CNRS-IN2P3 / INPG, \\
        53 Avenue des Martyrs, F-38026 Grenoble, France}
\ead{klasen@lpsc.in2p3.fr}

\begin{abstract}
\noindent
 We perform a NLO QCD analysis of single-diffractive dijet production in
 proton-antiproton collisions. By comparing the ratio of single- and
 non-diffractive cross sections to data from the Tevatron, the rapidity-gap
 survival probability is determined as a function of the momentum fraction of the
 parton in the antiproton. Assuming Regge factorization, this probability can be
 interpreted as a suppression factor for the diffractive structure function
 measured in deep-inelastic scattering at HERA. In contrast to the observations
 for photoproduction, the suppression factor in proton-antiproton collisions
 depends on the momentum fraction of the parton in the Pomeron even at NLO.
\end{abstract}


\end{frontmatter}

\def\d{{\rm d}}
\def\F{\tilde{F}}
\def\p{I\!\!P}
\def\R{\tilde{R}}

\def\lp{\left. }
\def\rp{\right. }
\def\lr{\left( }
\def\rr{\right) }
\def\le{\left[ }
\def\re{\right] }
\def\lg{\left\{ }
\def\rg{\right\} }
\def\lb{\left| }
\def\rb{\right| }

\def\beq{\begin{equation}}
\def\eeq{\end{equation}}
\def\bea{\begin{eqnarray}}
\def\eea{\end{eqnarray}}

%
\vspace*{-115mm}
\noindent {\rm LPSC 10-099}\\
\vspace*{ 106mm}
%

\section{Introduction}
\nin

In high-energy hadronic collisions, a large fraction of the events is produced
diffractively and contains one or more leading (anti-)protons or excited hadronic
states $Y$ with relatively low mass $M_Y$, which are separated from the central
hard process by rapidity gaps. Hard diffraction has been studied intensely in
deep-inelastic $ep$ scattering (DIS) at HERA, where it can be understood due to
the presence of the large squared photon virtuality $Q^2$ in terms of the Pomeron
structure function $F_2^{D}(x,Q^2,\xi,t)=\sum_if_{i/p}^D\otimes\sigma_{\gamma^*i}
$, factorizing into diffractive parton density functions (DPDFs) $f_{i/p}^D$ and
perturbatively calculable partonic cross sections $\sigma_{\gamma^*i}$
\cite{Collins:1997sr}. Assuming in addition Regge factorization, the DPDFs
$f_{i/p}^D=f_{i/\p}(\beta=x/\xi,Q^2)f_{\p/p}(\xi,t)$ can be parameterized in
terms of a Pomeron flux factor $f_{\p/p}$, which depends on the longitudinal
momentum fraction of the pomeron in the proton $\xi$ and the squared four-momentum
transfer at the proton vertex $t$, and PDFs in the Pomeron $f_{i/\p}$,
which depend on the momentum fraction of the partons in the pomeron $\beta$ and
an unphysical factorization scale $Q^2$, and fitted to
diffractive DIS data \cite{Aktas:2006hy}. Diffractive dijet production
data add further constraints, in particular on the gluon distribution
\cite{Aktas:2007bv}. At high $\xi$, a subleading Reggeon contribution must be
taken into account.

As $Q^2$ decreases, the applicability of perturbative QCD must be ensured by
sufficiently large jet transverse energies $E_{T1,2}$. From the absorption of
collinear initial-state singularities at next-to-leading order (NLO)
\cite{Klasen:2005dq}, the photon develops a hadronic structure and contributes
not only with direct, but also resolved processes \cite{Klasen:2004ct}. These
processes occur also in hadron-hadron scattering, where factorization is broken
due to the presence of soft interactions in both the initial and final states
\cite{Collins:1992cv}, and factorization breaking is indeed observed in dijet
photoproduction at NLO \cite{Klasen:2004qr}. Similar effects may also arise in
dijet production with a leading neutron \cite{Klasen:2001sg}. Soft rescattering
must be understood, if one wishes to exploit processes like diffractive Higgs
production, which offers a promising alternative to the difficult inclusive
diphoton signal in the low-mass region \cite{Royon:2003ng}. In this article, we
present the first NLO analysis of diffractive dijet production in hadron
collisions \cite{Klasen:2009bi}.

\section{Diffractive dijet production at the Tevatron}
\nin

Dijet production with a leading antiproton has been measured by the CDF
collaboration at the Tevatron in the range $|t|<1$ GeV$^2$, used also by the H1
collaboration for the extraction of their PDFs 2006 Fit A,B and 2007 Fit Jets, but
at slightly larger $\xi\in[0.035;0.095]$ \cite{Affolder:2000vb}. From the ratio
$\tilde{R}(x_{\bar{p}}=\beta \xi)$ of single-diffractive (SD) to
non-diffractive (ND) cross sections as a function of $x_{\bar{p}}=\sum_iE_{Ti}/
\sqrt{s}e^{-\eta_i}$, integrated over $E_{Ti}>7$ GeV and $|\eta_i|< 4.2$, an
effective diffractive structure function $\tilde{F}_{JJ}^D(\beta)=\tilde{R}\times
F_{JJ}^{ND}$ was extracted assuming $F_{JJ}^{ND}(x,Q^2)=x[f_{g/p}+4/9
\sum_if_{q_i/p}]$ with GRV 98 LO proton PDFs and compared to older H1 diffractive
structure functions at fixed $Q^2=75$ GeV$^2$. These were found to overestimate
the measured diffractive structure function by about an order of magnitude, while
the correction to be applied to the H1 measurements with $M_Y<1.6$ GeV, when
compared to the CDF measurements with leading antiprotons, should only be about
23\% \cite{Aktas:2006hy}.

In a subsequent publication, the CDF collaboration analyzed in a similar way data
at $\sqrt{s}=630$ GeV, not only 1800 GeV, in order to test factorization, Pomeron flux renormalization, and
rapidity-gap survival models \cite{Kaidalov:2001iz}, but found no significant
energy dependence \cite{Affolder:2001zn}. In addition they restricted the
$|t|$-range to below 0.2 GeV$^2$ and the average transverse jet energy to
$\bar{E}_T>10$ GeV, which has the advantage of removing the infrared sensitivity
of two equal cuts on $E_{T1,2}$ \cite{Klasen:1995xe}. In both publications, the
extraction of $\tilde{F}_{JJ}^D$ was performed on the basis of leading order (LO)
QCD and assuming that the $t$-channel gluon exchange in the partonic cross
sections dominates.

\section{NLO QCD analysis}
\nin

\begin{figure}
 \centering
 \includegraphics[width=.49\textwidth]{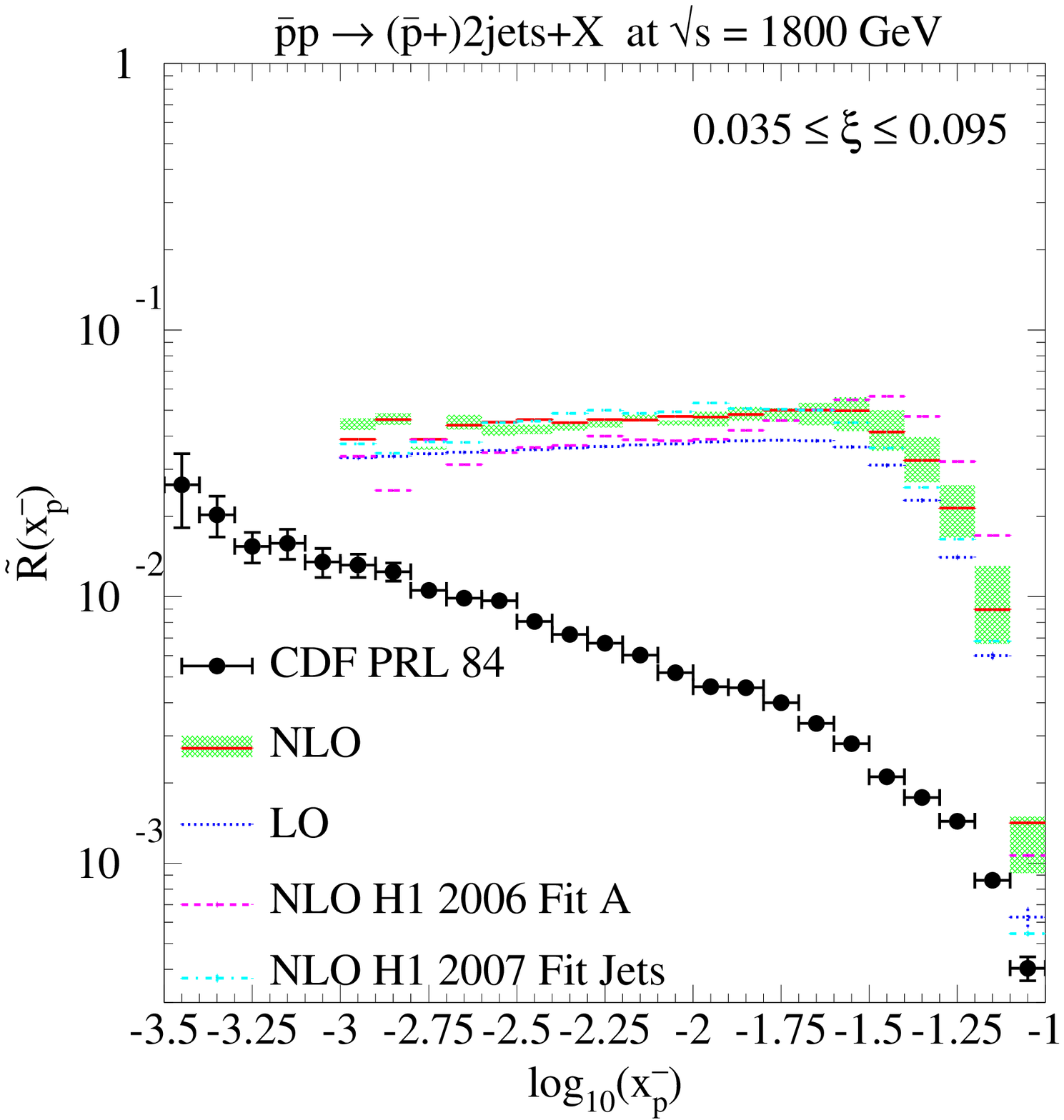}
 \includegraphics[width=.49\textwidth]{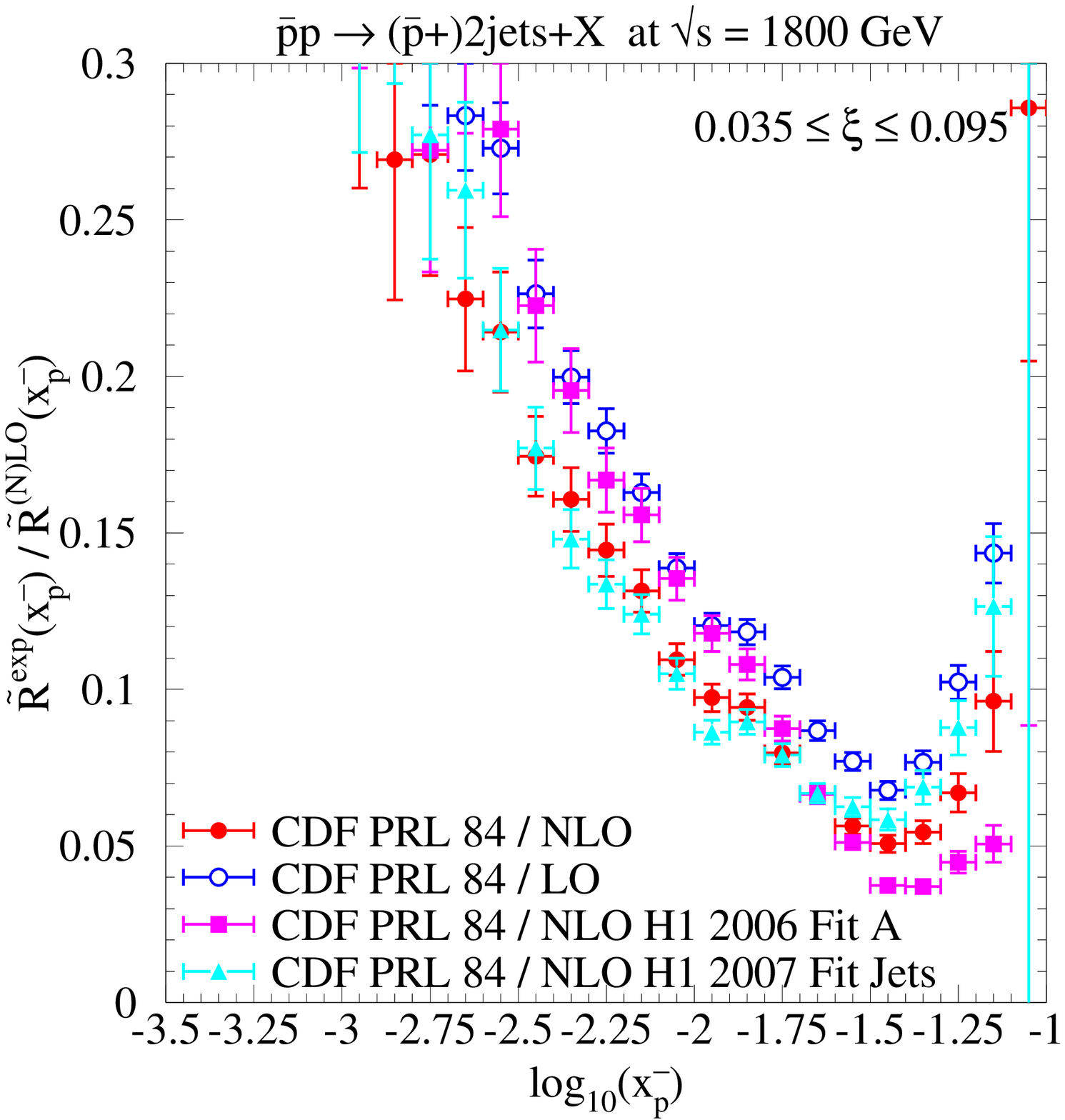}
 \caption{\scriptsize
 \label{fig:1}Left: Ratio $\R$ of SD to ND dijet cross sections
 as a function of the momentum fraction of the parton in the antiproton,
 computed at NLO (with three different DPDFs) and at LO and compared
 to the Tevatron Run I data from the CDF collaboration. Right: Double
 ratio of experimental over theoretical values of $\R$, equivalent to
 the factorization-breaking suppression factor required for an accurate
 theoretical description of the data (color online).}
\end{figure}

Our NLO QCD analysis is based on previous work on inclusive dijet photoproduction
\cite{Klasen:1996it}, where the resolved photon contribution can be directly
applied to dijet hadroproduction \cite{Klasen:1996yk}. It takes into account the
convolution of Pomeron flux factors and PDFs, modern CTEQ6.6M proton and H1
2006 and 2007 Pomeron PDFs with $Q^2=E_{T1}^2$ on an event-by-event basis, and the
complete set of parton-parton scattering cross sections. Jets are defined by a
cone with radius $R=0.7$ as in the CDF experiment and imposing a parton
separation of $R_{\rm sep}=1.3R$. The normalized control distributions $1/\sigma\,
\d\sigma/\d\bar{E}_T$ and $1/\sigma\,\d\sigma/\d\bar{\eta}$ (not shown) are in
good agreement with the SD and ND measurements, in particular for $E_{T2}>6.6$
GeV removing the infrared sensitivity in the first CDF analysis and for
$\sqrt{s}=630$ GeV in the second CDF analysis \cite{Klasen:2009bi}. However, the
ratio $\tilde{R}$ of SD to ND dijet cross sections as a function of $x_{\bar{p}}$
shown in Fig.\ \ref{fig:1} (left)
for $\sqrt{s}=1800$ TeV and $|t|<1$ GeV$^2$ significantly overestimates the data
for all employed diffractive PDFs and even more at NLO than at LO, due to an
average ratio of SD to ND $K$-factors of 1.35 (1.6 at $\sqrt{s}=630$ GeV).
Remember that all theoretical predictions should be divided by a factor of 1.23
to correct for diffractive dissociation included in the H1 DPDFs.

\begin{figure}
 \centering
 \includegraphics[width=0.49\textwidth]{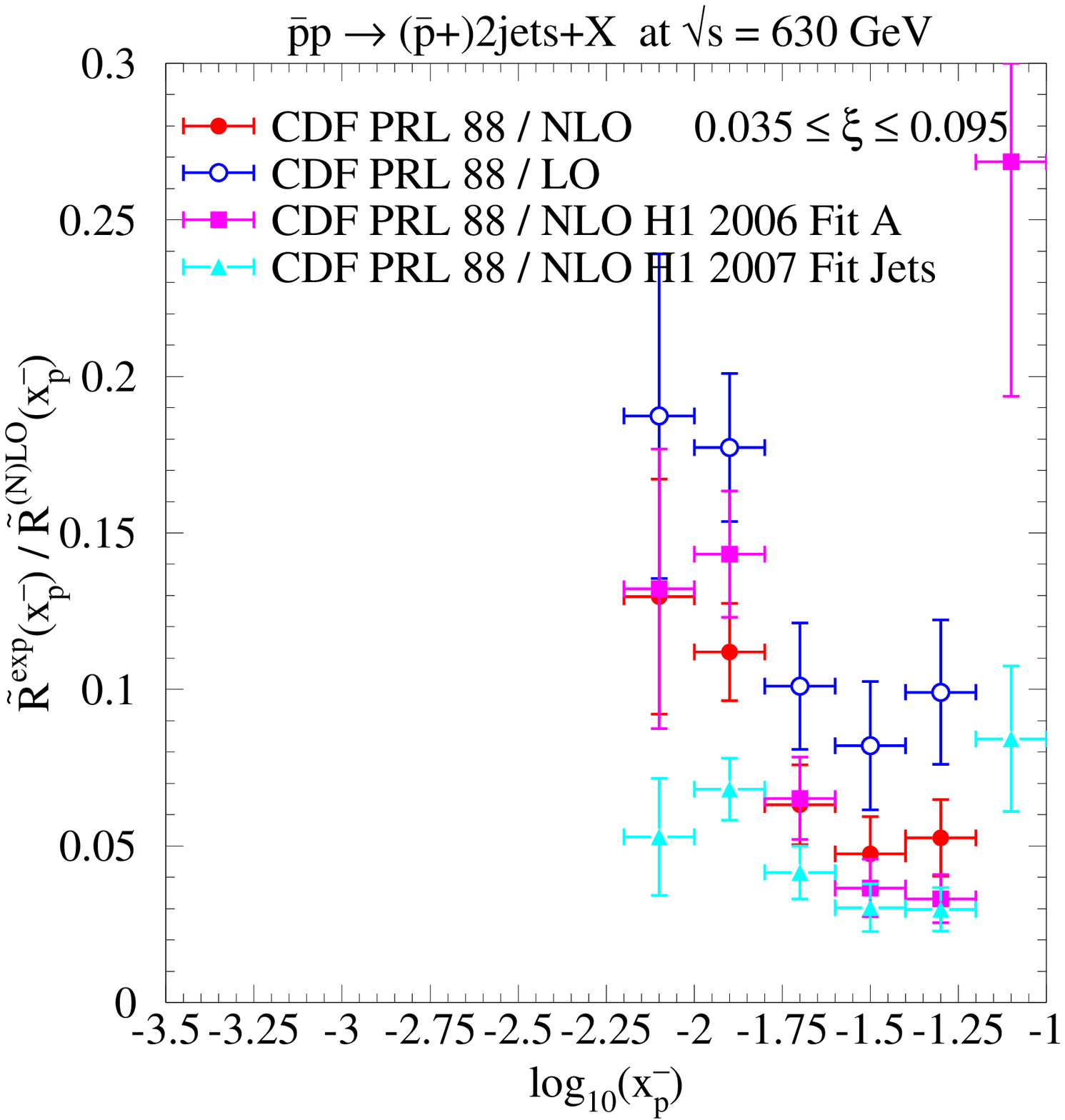}
 \includegraphics[width=0.49\textwidth]{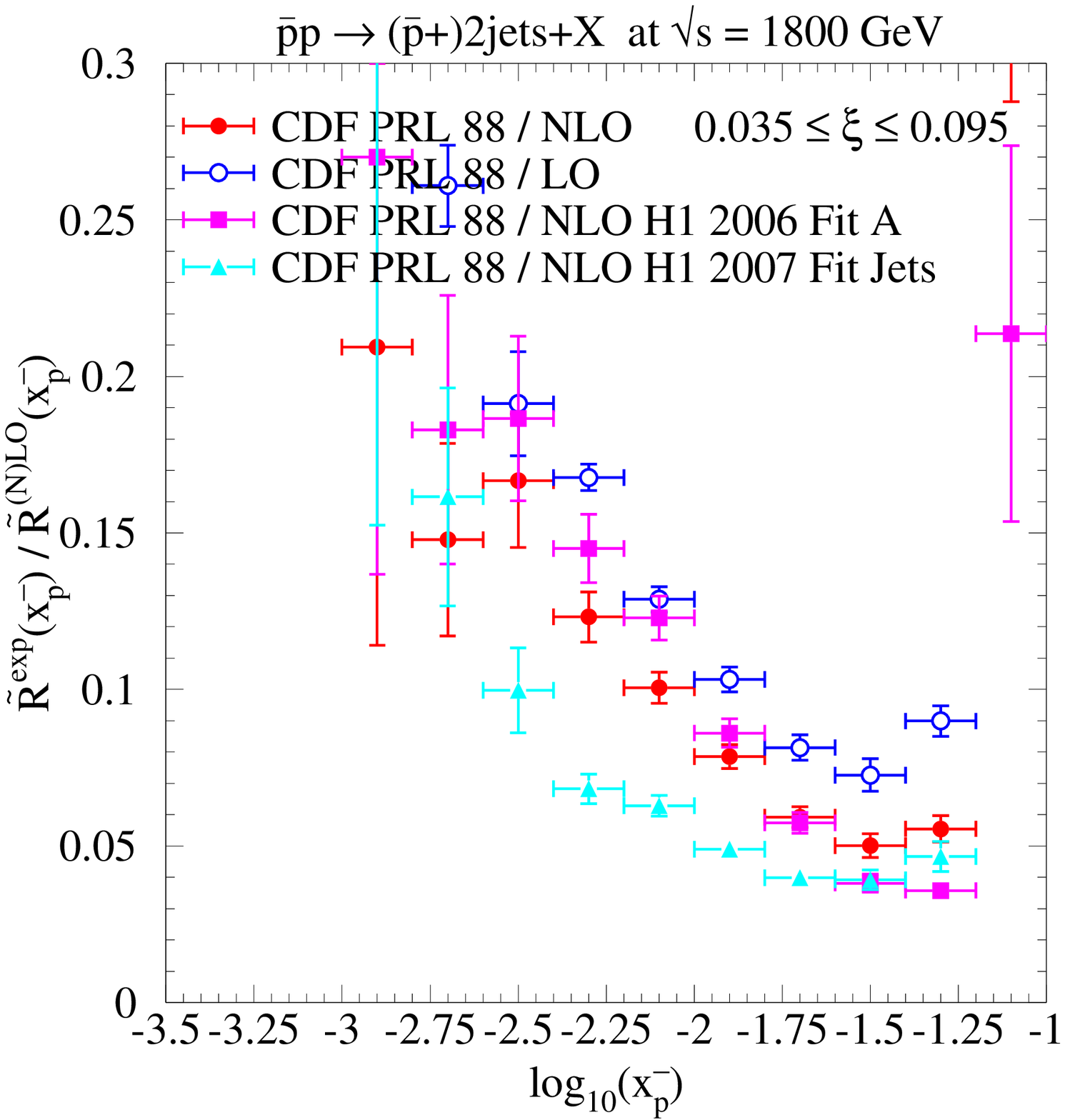}
 \caption{\scriptsize
 \label{fig:2}Double ratios of experimental over theoretical
 values of $\R$, equivalent to the factorization-breaking suppression
 factor required for an accurate theoretical description of the data
 from the Tevatron at $\sqrt{s}=630$ (left) and 1800 GeV (right)
 (color online).}
\end{figure}

The double ratios $\tilde{R}^{\rm exp}/\tilde{R}^{\rm (N)LO}$ in Fig.\ \ref{fig:1}
(right) and Fig.\ \ref{fig:2} can be interpreted as suppression factors or
rapidity-gap survival probabilities, which are smaller than one due to soft
rescattering. Their qualitative behavior in these three figures is very similar:
we observe a dependence on $x_{\bar{p}}$ with a minimum at $\log_{10}(x_{\bar{p}})
\simeq -1.5$ $(x_{\bar{p}}\simeq 0.032)$, a rise towards smaller $x_{\bar{p}}$ by
up to a factor of five, a smaller rise towards larger $x_{\bar{p}}$, and an
appreciable dependence of the suppression factor on the chosen diffractive PDFs.
The corresponding suppression factors for $\tilde{F}_{JJ}^D(\beta)$ (not shown),
computed with the same assumptions on $F_{JJ}^{ND}$ as those taken by CDF, are
minimal with the value $\sim 0.05$ at $\beta = 0.5$ and rise to $\sim 0.12$ for
$\sqrt{s}=1800$ GeV and to $\sim 0.1$ for $\sqrt{s}=630$ GeV at $\beta=0.1$ and
with H1 2006 Fit B. Here, the effect of NLO corrections is partially compensated
by the simplifications inherent in the calculation of $\tilde{F}_{JJ}^D$.
The strong rise of the suppression factors to values of $\sim 0.25$ for
$\beta<0.05$ $[\log_{10}(x_{\bar{p}})<-2.5]$ at $\sqrt{s}=1800$ GeV may not be
conclusive, since the H1 DPDFs are largely unconstrained in this region. On the
other hand, the rise at $\log_{10}(x_{\bar{p}})>-1.5$ can be attributed to the
Reggeon contribution, that should be added at large $\xi$.  Note also that the
variation of the suppression factor is considerably reduced for the H1 2007 Fit
Jets, which has been obtained with additional constraints from  diffractive DIS
dijet production. 

The extraction of $\tilde{F}^D_{JJ}(\beta)$ from $\tilde{R}(x_{\bar{p}})$ is
based on the assumption that the latter is only weakly $\xi$-dependent and can be
evaluated at an average value of $\bar{\xi}=0.0631$. This weak $\xi$-dependence
is indeed observed in the newer CDF data and also in our theoretical calculations,
which reflect the $\xi$-dependence of the H1 fits to the Pomeron flux factors
$f_{\p/\bar{p}}(\xi,t)\propto \xi^{-m}$ with  $m\simeq 1.1$ (0.9 in the CDF fit
to their data). The (small) difference of the theoretical (1.1) and experimental
(0.9) values of $m$ can be explained by a subleading Reggeon contribution, which
has not been included in our predictions. To study its importance, we have
computed the ratio of the Reggeon over the Pomeron contribution to the LO
single-diffractive cross section at $\sqrt{s}=1800$ GeV. The Reggeon flux factor
was obtained from H1 2006 Fit B and convolved, as it was done in this fit, with
parton densities in the pion \cite{Owens:1984zj}. Very similar results were
obtained for the H1 2007 Fit Jets Reggeon flux. On average, the Reggeon
adds a 5\% contribution to the single-diffractive cross section, which is
smaller at small $x_{\bar{p}}=\xi\beta$
and $\xi$ (2.5\%) than at large $x_{\bar{p}}$ and $\xi$ (8\%). This corresponds to
the graphs shown in Figs.\ 5 ($\xi=0.01$) and 6 ($\xi=0.03$) of Ref.\ 
\cite{Aktas:2006hy}, e.g.\ at $Q^2=90$ GeV$^2$. While the Reggeon contribution
thus increases the diffractive cross section and reduces the suppression factor
at large $x_{\bar{p}}$, making the latter more constant, the same is less true at
small values of $x_{\bar{p}}$.

Any model calculation of the suppression factor or rapidity-gap survival
probability must try to explain two points, first the amount of suppression,
which is $\sim 0.1$ at $\beta=0.1$, and second its dependence on the variable
$\beta$ (or $x_{\bar{p}}$). Such a calculation has been performed by Kaidalov et
al.\ \cite{Kaidalov:2001iz}. In this calculation, the hard scattering cross
section for the diffractive production of
dijets was supplemented by screening or absorptive corrections on the basis of
eikonal corrections in impact parameter space. The parameters of the
eikonal were obtained from a two-channel description of high-energy inelastic
diffraction. The exponentiation of the eikonal stands for the exchange of
multi-Pomeron contributions, which violate Regge and QCD factorization and modify
the predictions based on single Pomeron and/or Regge exchange. The obtained
suppression factor is not universal, but depends on the details of the hard
subprocess as well as on the kinematic configurations. The first important
observation from this calculation is that in the Tevatron dijet analysis the mass
squared of the produced dijet system $M_{JJ}^2=
x_p\beta \xi s$ as well as $\xi$ are almost constant, so that small $\beta$
implies large $x_p$.
The second important ingredient in this calculation is the assumption that the
absorption cross section of the valence and the sea components,
where the latter includes the gluon, of the incoming proton are different, in
particular, that the valence and sea components correspond to smaller and larger
absorption. For large $x_p$ or small $\beta$, the valence quark contribution
dominates, which produces smaller absorptive cross sections as compared to the
sea quark and gluon contributions, which dominate at small $x_p$. Hence the
survival probability increases as $x_p$ increases and $\beta$ decreases. The
convolution of the $\beta$-dependent absorption corrections with older H1 DPDFs
\cite{H1} led to a prediction for $F^D_{JJ}(\beta)$, which was in very good
agreement with the corresponding experimental distribution (see Fig.\ 4 in
\cite{Kaidalov:2001iz}). A similar correction for soft rescattering of our
single-diffractive NLO cross sections based on the more recent DPDFs of H1
should lead to a very similar result.
An alternative model for the calculation of the suppression factor was developed
by Gotsman et al.\ \cite{G}. However, these authors did not convolve their
suppression mechanism with the hard scattering cross section. Therefore a direct
comparison to the CDF data is not possible.

At variance with the above discussion of diffractive dijet production in
hadron-hadron scattering, the survival probability in diffractive dijet
photoproduction was found to be larger ($\sim0.5$ for global suppression,
$\sim0.3$ for resolved photon suppression only) and fairly independent of
$\beta$ \cite{Klasen:2004qr,Aktas:2007hn}. This can be explained by the fact that
the HERA analyses are restricted to large values of $x_\gamma\geq 0.1$ (as opposed
to small and intermediate values of $x_p=0.02$ ... $0.2$ at the Tevatron), where
direct photons or their fluctuations into perturbative or vector meson-like
valence quarks dominate. The larger suppression factor in photoproduction
corresponds also to the smaller center-of-mass energy available at HERA.

\section*{Acknowledgements}
\nin
It is a pleasure to thank G.\ Kramer for his collaboration and S.\ Narison for
the invitation to this conference.

\end{document}